# COMPARATIVE INVESTIGATION FOR ENERGY CONSUMPTION OF DIFFERENT CHIPSETS BASED ON SCHEDULING FOR WIRELESS SENSOR NETWORKS

Monica[1] and Ajay K Sharma[2]

Department of Computer Science and Engineering,
Dr B R Ambedkar National Institute of Technology, Jalandhar, Punjab
pikhen.monica@gmail.com[1], sharmaajayk@nitj.ac.in[2]

## ABSTRACT

*Rapid progress in microelectromechanical system (MEMS) and radio frequency (RF) design has enabled the development of low-power, inexpensive, and network-enabled microsensors. These sensor nodes are capable of capturing various physical information, such as temperature, pressure, motion of an object, etc as well as mapping such physical characteristics of the environment to quantitative measurements. A typical wireless sensor network (WSN) consists of hundreds to thousands of such sensor nodes linked by a wireless medium. In this paper, we present a comparative investigation of energy consumption for few commercially available chipsets such as TR1001, CC1000 and CC1010 based on different scheduling methods for two types of deployment strategies. We conducted our experiment within the OMNeT++ simulator.*

## 1. INTRODUCTION

The progress of hardware technology in low-cost, low-power, small-sized processors, transceivers, and sensors has facilitated the development of wireless sensor networks. A distributed sensor network is a self-organized system composed of a large number (hundreds or thousands) of low-cost sensor nodes. Self-organization means that the system can achieve the necessary organizational structures without requiring human intervention, i.e., the sensor network should be able to carry out functional operations through cooperation among individual nodes rather than set up and operated by human operators [1]. Sensor nodes are usually battery based, with limited energy resources and capabilities; it is difficult or unpractical to recharge each node. The far-ranging potential applications of sensor networks include: (1) system and space monitoring; (2) habitat monitoring; (3) target detection and tracking; and (4) biomedical applications.

Sensor networks are subject to tight communication and computation constraints. These constraints are due to size and cost limitations: many of the most interesting applications require tiny and cheap sensors [2]. This creates an interesting interplay between energy (and device size) spent on communication versus computation, and it is unclear which of the two is the true bottleneck to use. In this paper we concentrate on the energy spent in communication as energy is vital and scarce resource that should be wisely managed in order to extend the lifetime of the network.

There are four major sources of energy waste in WSNs: retransmission due to collision or congestion, idle channel sensing for the packets that has never been sent, overhearing the packets that are destined to other nodes and control packet overhead [3]. To alleviate these energy wastes, several MAC protocols for wireless sensor networks have been proposed. These protocols are classified into two types: scheduling based and contention based. Scheduling based protocols are usually TDMA-based protocols [4, 5] in which each sensor node is assigned one of time slots and can communicate only in the assigned time slot. In contention based or random access users access the network based only on its local information, e.g. the state of its queue or the state of the channel that it senses, which is relatively random for other users in the network. CSMA [6, 7] is one of the most popular choices for random access networks due to its simple and effective design [8].

The MERLIN (Mac Energy Efficient, Routing and Location Integrated) protocol for energy management within the WSN, offers an approach to such problems of wasteful energy consumption. The MERLIN [9] protocol divides the sensor networks into time zones. It uses a combination of TDMA and CSMA techniques to decrease the node activity and the number of collisions as well as to increase the scalability. MERLIN adopts multicast upstream and multicast downstream transmission to relay information both to the gateway and away from it. Time zone updates and exchange of synchronization packets happens through periodic local broadcast. Each node locally stores a scheduling table that regulates its periodic activity. We have implemented MERLIN protocol in order to obtain our desired results.

This paper is structured as follows: Section 2 discusses the related work in the area of energy consumption based on scheduling methods and deployment strategies. Section 3 describes the scheduling process and discusses the six scheduling methods. Section 4 gives a brief description of the deployment strategies. In Section 5 we provide a detailed simulation network setup and simulation results which compares the six scheduling methods based on energy consumption. Concluding remarks and directions for the future research can be found in Section 6.

## 2. RELATED WORK

A large number of scheduling algorithms have been proposed to reduce the energy consumption at all levels of the wireless sensor networks. In 2004, Barbara Hohlt et al., [10] proposed a distributed on-demand power-management protocol for collecting data in sensor networks. In the paper, they presented a dynamic distributed time division scheduling protocol that facilitated power management by enabling nodes to turn off their radios during idle slots. Implementation and simulation have both shown that power scheduling reduces the energy consumption at all levels of the network and that the network can adapt schedules locally to changing demand. The paper also showed that non-idle slots can be significantly reduced by locally managing supply and demand.

A.G. Ruzzelli et al., in 2005 [11], has investigated the use of intelligent agents in the delivery of adaptivity at the networking layers. This research has described a method of optimizing energy resources in times when unexpected or heavy network activity occurs. Three instruments facilitate this: the provision of two efficient and interchangeable scheduling tables; the ability to generate virtual network sectors; the adoption of autonomous mobile agents.

Antonio G. Ruzzelli et al., in 2005 [12], addressed the tradeoff between energy conservation and latency. They presented an experimental evaluation of two node scheduling regimes within MERLIN (Mac Energy efficient, Routing and Localization INtegrated), an energy-efficient low-latency integrated protocol for WSNs. In particular they contrasted the X and V scheduling family schemes with respect to the following properties: network setup time, network lifetime and message latency.

Sinem Coleri Ergen and Pravin Varaiya in 2005 [13] proposed TDMA Scheduling algorithm for sensor networks. Two centralized heuristic algorithms for solving the problem: One is based on direct scheduling of the nodes, node based scheduling, whereas the other is based on scheduling the levels in the routing tree before scheduling the nodes, level based scheduling were proposed. The performance of these algorithms depends on the distribution of the nodes across the levels.

Yunali Wang et al., in 2006 [14] reported the problem of efficiently link scheduling algorithm in wireless sensor networks with interference. Shubham Jain and Sanjay Srivastava in 2007 [15], presented a summary, comparative discussion and classification of the selected eight algorithms on the distributed scheduling in sensor networks Nikolaos A. Pantazis et al., in 2008 [16] proposed TDMA based scheduling scheme that balances energy saving and end-to-end delay. This balance could be achieved by an appropriate scheduling of wakeup intervals. It allowed data packets to be delayed by only one sleep interval for the end-to-end transmission from the sensors to the gateway. The proposed scheme achieves the reduction of the end-to-end delay caused by the sleep mode operation while at the same time it maximizes the energy savings.

In 2005, Ali Iranli et al., [17] investigated and developed energy sufficient strategies for deployment of WSN for the purpose of monitoring some phenomenon of internet in the coverage region. In this paper, the operational advantages of two level hierarchical architecture over flat network architecture have been studied. The paper formulated and solved the problem of assigning positions and initial energy levels to micro-servers and concurrently partitioning the sensors into clusters assigned to individual micro-servers so as to maximize the monitoring lifetime of the two-level WSN subject to total energy budget.

## 3. SCHEDULING FOR WSNs

Literature has proposed various scheduling algorithms that maximize network performance while minimizing energy usage. It has been observed that the efficiency of the MAC design is crucial for sensor networks due to the scarce energy and bandwidth resources of these systems. Based on the assumption that the users in the system are transmitting independent data and that they are competing for the use of the transmission channel, MAC protocols are designed to allocate separate interference-free channels to each user. This is typically achieved through either *random access* or *deterministic scheduling*. The purpose of scheduling is to allocate time zone slots. Nodes in the same time-zone use the same slot to transmit. Nodes send only in one slot per frame that slot being determined by the node's hop count. The timing of the slots prevents most collisions. Antonio G. Ruzzelli et al. [12] have contrasted the X and V scheduling family scheme. The comparison of more scheduling methods with the earlier ones is a logical continuation of this research.

All nodes share a common sensing task; hence there is a sensing redundancy. This employs that not all the sensors are required to perform the sensing task during the whole system lifetime. So it is important to schedule the nodes. The concept of scheduling comes from the operating systems. Round Robin is one of the simplest scheduling algorithms for processes in the operating systems, which assigns time slices to each process in equal portions and in order, handling all processes without priority. In networks processes may be replaced by packets or nodes [16].

A proper schedule not only avoids collisions by silencing the interferes of every receiver node in each time slot but also minimizes the number of time slots hence the latency. The larger latency may require a higher data rate and hence higher energy consumption [17]. The aim of the scheduling is to allocate the timeslots to the network nodes in order to assign them the periods of activity and inactivity. Scheduling helps in synchronization of the neighboring nodes for the transmission and reception of packets. Scheduling follows a particular data pattern. Typically there can be two types of data patterns: continuous and discontinuous. In continuous data pattern, a packet is forwarded from source to destination without interruption while discontinuous data pattern introduce some delay in forwarding process. This delay can be used to avoid collisions or save energy. An appropriate scheduling policy can be developed according to such expected data patterns [18].

MERLIN adopts a time zone based transmission scheduling which allows the nodes in the same time zone to use the same slot to transmit. The scheduling table is usually transmitted by the gateway during the initialization phase, and schedules the periods of nodes activity and inactivity.

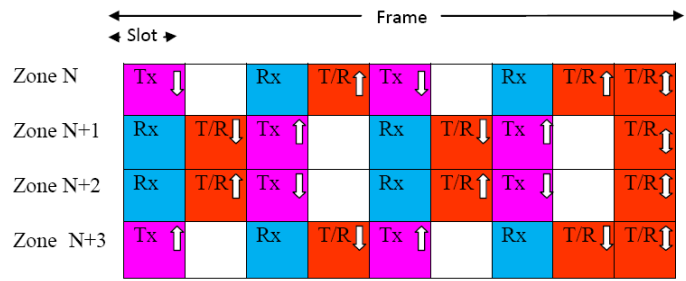

(a)

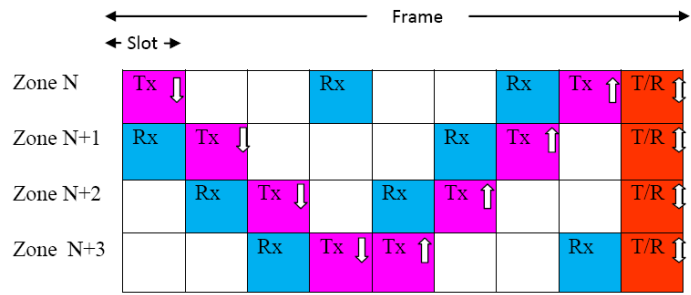

(b)

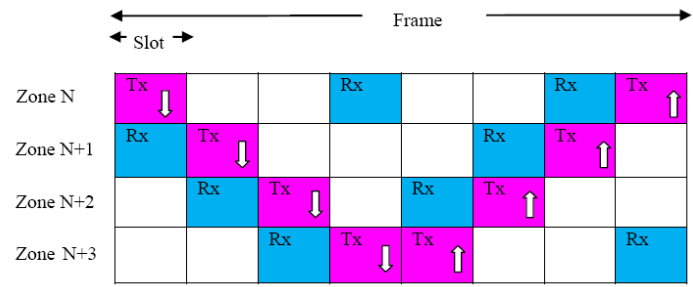

(c)

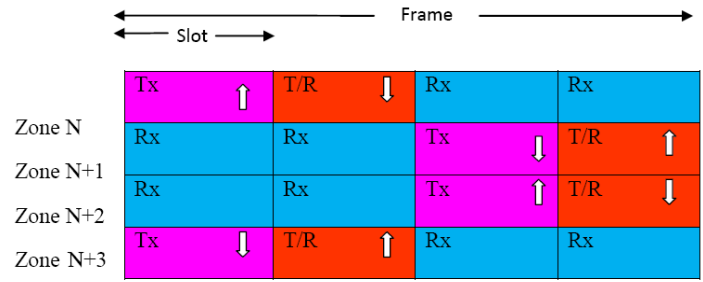

(d)

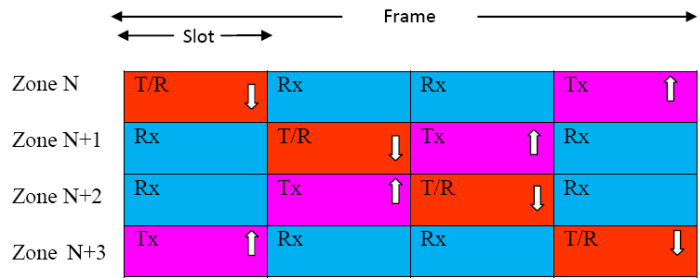

(e)

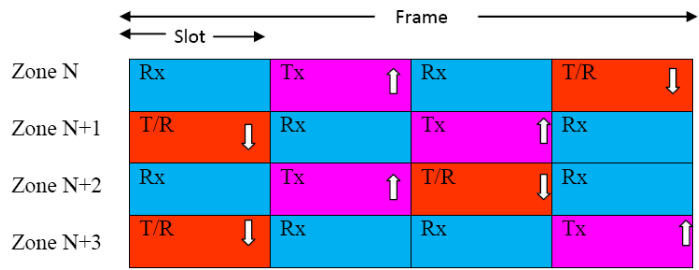

(f)

Figure 1: Scheduling methods (a) X scheduling, (b) V (9x9) scheduling, (c) V scheduling, (d) Leon (4x4) Crossed Shifted scheduling, (e) Leon (4x4) Crossed Not Shifted scheduling, and (f) (4x4) Crossed Shifted scheduling.

Figure 1(a) and 1(c) shows the two different types of scheduling methods investigated for MERLIN. These are referred to as X and V Scheduling due to the X/V-shape of the communication flow. Many other scheduling types are derived from these two basic types of scheduling. Variation of either slot length or frame time causes different delay in packet scheduling. The total length of the table is equal to the length of a single frame while each small rectangle represents a time slot. In the tables, the nodes within the same time zone contend the channel for transmission, the adjacent zone owns the slot for reception and the nodes in the further time zones are in sleep mode. The tables represent the models of inter-zone scheduling that provide different priorities to energy consumption.

In X scheduling, nodes remain active most of the time and hence higher channel utilization is achieved here. They can also hold concurrent transmission slots between nearby time zone. Scheduling of further zones can achieved by appending the same table. X table performs the upstream and downstream concurrent transmission by forwarding a packet to 8 time zones towards the gateway or in opposite direction within the same frametime. X table allocates 8 timeslots for upstream, 8 timeslots for downstream and 1 timeslot for local broadcast. In V (9x9) scheduling as can be seen in Figure 1 (b), half of the slots are allocated to downstream data traffic and rest half are for the upstream data traffic. In this way the downstream and upstream data traffic does not intersect. The nodes have more inactive slots. The last slot is allocated to local broadcast. The V (9x9) table performs upstream and downstream non-interfering transmission by forwarding a packet to 4 time zones towards the gateway or in the opposite direction within the same frame time.

Figure 1 (c) shows the V scheduling. In this scheduling we have not considered a dedicated slot for local broadcast. The frame is divided into 8 slots. Half of the slots are allocated to downstream data traffic and remaining half slots are allocated to upstream data traffic. This division of traffic allocation helps avoiding interference between the upstream and downstream data traffic. The Figure 1 (d) shows the Leon (4x4) Crossed Shifted scheduling. In this scheduling, a frame has been divided into 4 slots. Transmission, reception and the local broadcast are done simultaneously. In this scheduling, as soon as the node finishes transmission or reception of the packet it goes into sleep mode. It performs the concurrent upstream and downstream data transmission.

Figure 1 (e) shows the Leon (4x4) Crossed Not Shifted scheduling. This scheduling divides the frame into 4 slots. In this scheduling, downward transmission is done by the local broadcast from the gateway. Upward and downward data traffic is transmitted diagonally. The Leon (4x4) Crossed Not Shifted scheduling performs upstream and downstream interfering transmission by forwarding a packet to 4 time zones towards the gateway or in the opposite direction within the same frame time. Figure 1 (f) shows (4x4) Crossed Shifted scheduling. In this scheduling, upward and downward data traffic can be scheduled simultaneously for different zones. Nodes in alternative zones can transmit simultaneously. This scheduling also divides the frame into 4 slots.

## 4. DEPLOYMENT CONSIDERATION

Deploying sensors to provide complete area coverage is another essential design problem in many WSN applications. Mainly three alternative deployment approaches have been proposed in literature. One among them is application-specific deterministic deployment, another is random deployment and the third one is grid based (also known as pattern-based) deployment [19, 20]. In deterministic deployment, the sensor nodes are placed deliberately in the required region. This type of deployment is suitable only for small-scale applications. Non-deterministic deployment is scalable to large-scale applications or hostile environments. In this type of deployment, the sensor nodes are thrown randomly to form a WSN. However, it could be very expensive since excess redundancy is required to overcome uncertainty. Grid-based deployment is an attractive approach for moderate to large-scale coverage-oriented deployment due to its simplicity and scalability. In this research, we focus on the difference in of energy consumption for the few commercially used sensor nodes when they are deployed in random and grid fashion.

### 4.1 Random deployment

Here, we considered random deployment of sensors. It means that the place of sensors in the network is not considered previously. Random deployment of sensor nodes in the physical environment may take several forms. It may be a one-time activity where the installation and use of a sensor network are strictly separate actions. Or, it may be a continuous process, with more nodes being deployed at any time during the use of the network; for example, to replace failed nodes or to improve the coverage area at certain locations [21, 22]. Randomized sensor deployment is quite challenging in some respects, since there is no way to configure a priori the exact location of each device. Additional post-deployment self-configuration mechanisms are required to obtain the desired coverage and connectivity. In case of a uniform random deployment, the only parameters that can be controlled a priori are the numbers of nodes and some related settings on these nodes, such as their transmission range. Figure 2 (a) shows the random deployment of sensor nodes for WSN.

## 4.2 Grid Deployment

There are three types of grid-based deployment corresponding to three regular shapes which can tile a plane without holes, namely, hexagon, square and equilateral triangle. Grid-based deployment is conducted by dropping sensors row-by row using a moving carrier. The time interval between consecutive droppings is controlled to achieve the desired distance. However, often this ideal deployment is not realistic due to placement errors [23]. In the unreliable sensor grid model, n nodes are placed on a square grid within a unit area, with a certain probability that a node is active (not failed), and a defined transmission range of each node.

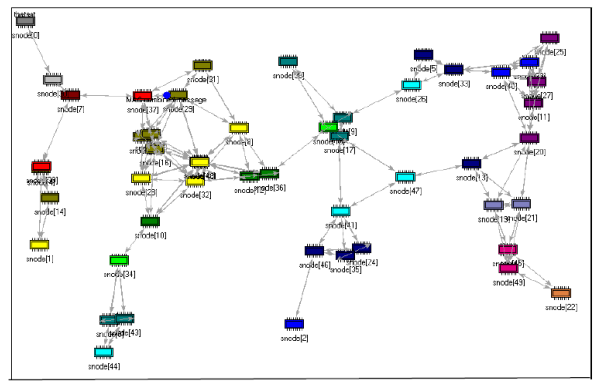

(a)

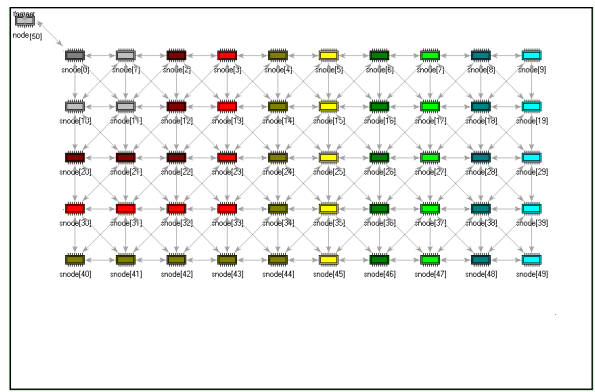

(b)

180

Figure 2: Deployment strategies for sensor nodes in wireless senor network (a) Random
(b) Grid

Adding nodes to ensure wireless connectivity is a challenging issue, particularly when there are location constraints in a given environment that dictate where nodes can or cannot be placed. If the number of available nodes is small with respect to the size of the operational area and required coverage, a balance between sensing and routing nodes has to be optimized. Figure 2 (b) shows the grid deployment of sensor nodes for WSN.

## 5. Experimentation and Results

In our simulation, we use a rectangular space of 800 x 500 centimeters as the sensing area, one of the corners being fixed at origin. We have chosen to work with two different scenarios, one with random deployment of sensor nodes and another with grid deployment of sensor nodes. Each of the scenarios consists of 50 sensor nodes including one gateway. The node transmitting range allows communication with direct neighbors only.

The simulation is aimed to evaluate the energy consumed by the sensor node for a particular type of deployment strategy with six different scheduling methods over varying payload. The Payload is the data sent from source node to the destination node. The packet length varies depending on the data carried with a payload that may vary from 0 to 127 bytes. Each graph has been obtained by running 7 independent simulations of 20 minutes each.

As the simulation begins, the sensor nodes are generated which gets randomly distributed over the sensing area. The nodes setup their connections with the neighbors and routing paths are established according to Multi Hop (MH) Routing Algorithm applied to sensor networks. All the simulations have been carried in OMNeT++ [24] discrete event simulator. The results are collected by applying all the six types of scheduling methods to a particular chipset deployed in one of the two deployment strategies i.e. Random or Grid Deployment over the increasing extra payload. We collected the results for three different chipsets i.e. TR1001, CC1000 and CC1010.

Figure 3 (a) shows a comparison of total energy consumed by the chipsets TR1001, CC1000 and CC1010 for V scheduling at (a) random deployment and (b) grid deployment. It is evident that in both the cases, TR1001 consumes lowest energy with a value of 0.067531 J and 0.213444 J for random and grid deployment respectively at the extra payload of 1023 bits. It is also concluded from the figure that TR1001 consumes less energy in case random deployment.

Figure 3 also shows that CC1000 consumes an energy of 0.124197 J and 0.529815 J at extra payload of 1023 bits, respectively for random and grid deployment. It can be analyzed that CC1000 chipset consumes less energy in case of random deployment than grid deployment. The Figure 3 indicates that CC1010 consumes 0.276794 J and 1.289059J of energy for random and grid deployment respectively at extra payload of 1023 bits. It can be observed that energy consumed by CC1010 in case of V scheduling is much lower for the random deployment.

It has been observed that there is a significant difference in energy consumption for the two deployments for the three chipsets in the case of V scheduling. It can be noticed that the energy consumed in case of random deployment is less than the energy consumed in the case of grid deployment for all the three chipsets.

Figure 4 shows a comparison of total energy for the three chipsets TR1001, CC1000 and CC1010 for V (9x9) scheduling for (a) random deployment and (b) grid deployment. It is observed that energy consumed by TR1001 is lowest among the other chipsets for both types of deployments. TR1001 consumes 0.195657 J and 0.180813 J of energy at the extra payload of 1023 bits for random and grid deployment respectively. It is also concluded that TR1001 consumes less energy in case of grid deployment.

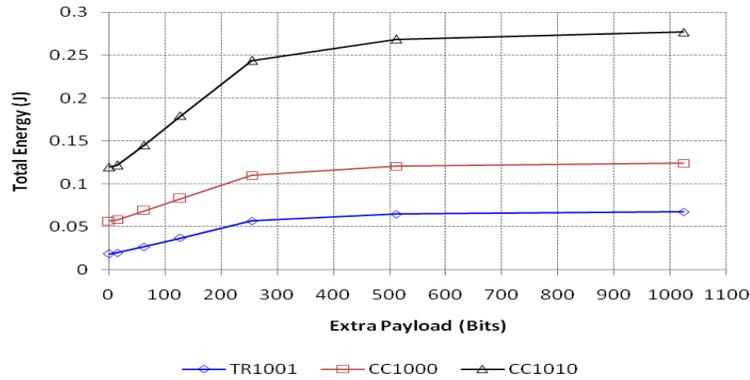

(a)

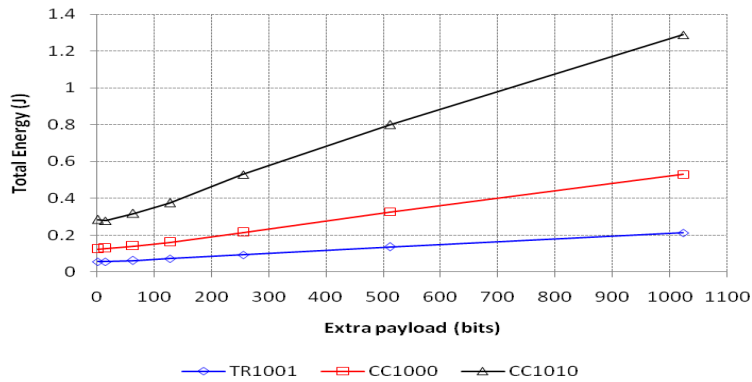

(b)

Figure 3: Comparison of Total Energy for the chipsets TR1001, CC1000 and CC1010 for V scheduling for (a) Random deployment, (b) Grid deployment.

As can be seen from Figure 4, chipset CC1000 consumes an energy of 0.341697 J and 0.340676 J at extra payload of 1023 bits, for random and grid deployment respectively. It can be analyzed that CC1000 chipset consumes less energy in case of grid deployment. The Figure 4 indicates that CC1010 consumes 0.817593 J and 0.819736 J of energy for random and grid deployment respectively at extra payload of 1023 bits. It can be observed that energy consumed by CC1010 in case of random deployment is lower than grid deployment.

It is also indicated from figure, that there is not much deviation in energy consumption by the three chips in both the random and grid scheduling. Hence it can be concluded that V (9x9) scheduling does not present a significant difference in energy consumption for the two deployments

Figure 5 shows a comparison of total energy consumed by the three chipsets TR1001, CC1000 and CC1010 for X scheduling for (a) random deployment and (b) grid deployment. It is observed that energy consumed by TR1001 is lowest among the other chipsets for both types of deployments. TR1001 consumes 0.300801 J and 0.168999 J of energy at the extra payload of 1023 bits for random and grid deployment respectively. It is also concluded from the figure that TR1001 consumes less energy in case grid deployment.

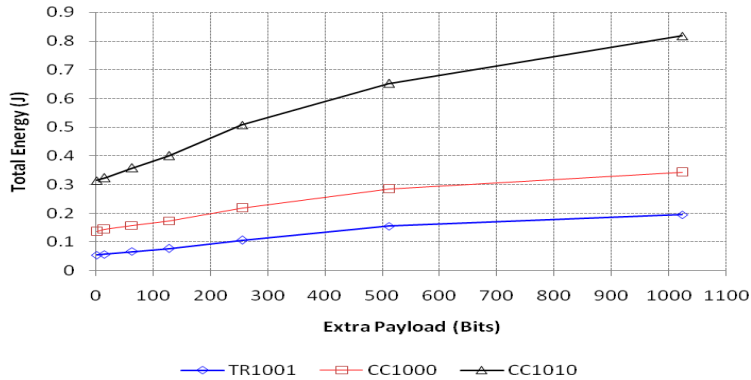

(a)

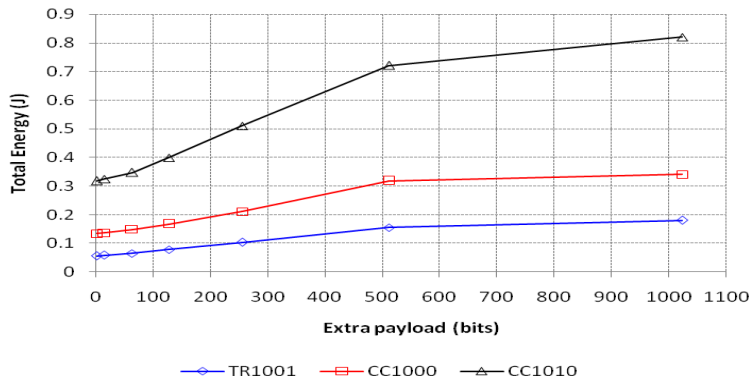

(b)

Figure 4: Comparison of Total Energy for the chipsets TR1001, CC1000 and CC1010 for the V (9x9) scheduling for (a) Random Deployment, (b) Grid deployment.

As evident from the Figure 5, CC1000 consumes an energy of 0.0.597699 J and 0.408986 J at extra payload of 1023 bits, for random and grid deployment respectively. It can be analyzed that CC1000 chipset consumes less energy in case of grid deployment. The Figure 5 indicates that CC1010 consumes 1.44315 J and 0.974794 J of energy for random and grid deployment respectively at extra payload of 1023 bits. It can be observed that energy consumed by CC1010 in case of grid deployment is lower than random deployment. The Figure 5 indicates that the two deployments show a significant difference in energy consumption for the X scheduling. Also it can be seen that TR1001 consumes minimum energy and CC1010 consumes maximum energy for both the deployments.

Figure 6 shows a comparison of total energy consumed by the three chipsets TR1001, CC1000 and CC1010 for Leon (4x4) Crossed Not Shifted scheduling for (a) random deployment and (b) grid deployment. TR1001 consumes 0.314237 J and 0.270063 J of energy at the extra payload of 1023 bits for random and grid deployment respectively. It is also observed that TR1001 consumes less energy in case grid deployment.

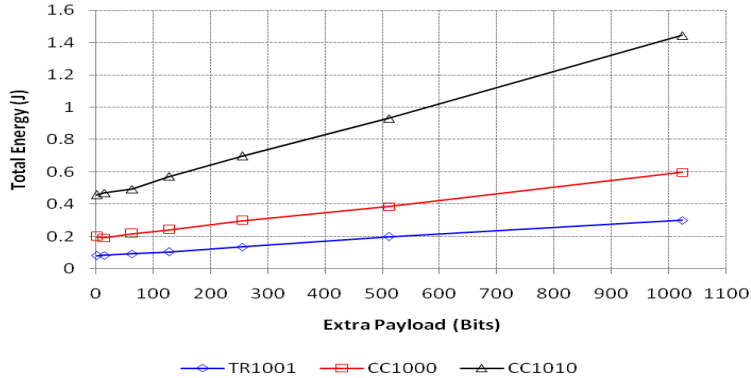

(a)

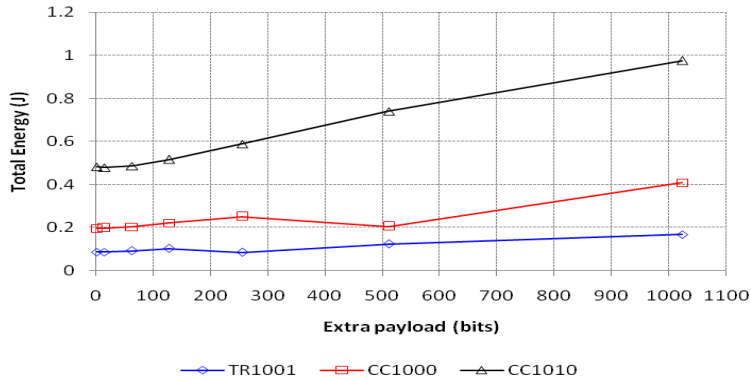

(b)

Figure 5: Comparison of Total Energy consumed by the chipsets TR1001, CC1000 and CC1010 for the X scheduling for (a) Random Deployment, (b) Grid deployment.

Figure 6 depicts that CC1000 consumes an energy of 0.529035 J and 0.530603 J at extra payload of 1023 bits, for random and grid deployment respectively. It can be analyzed that CC1000 chipset consumes less energy in case of random deployment. The Figure 6 indicates that CC1010 consumes 1.289097 J and 1.288425 J of energy for random and grid deployment respectively at extra payload of 1023 bits. It can be observed that energy consumed by CC1010 in case of grid deployment is lower than random deployment. It is also observed that, Leon (4x4) Crossed Not Shifted scheduling does not show a remarkable difference in energy consumptions for the two types of deployment. Also in both the cases, TR1001 consumes

184

minimum energy and CC1010 consumes maximum energy. It is also evident from the figure that with the increase in extra payload, the energy consumption also increases. This is due to the fact that energy consumption is directly proportional to the payload of transmission.

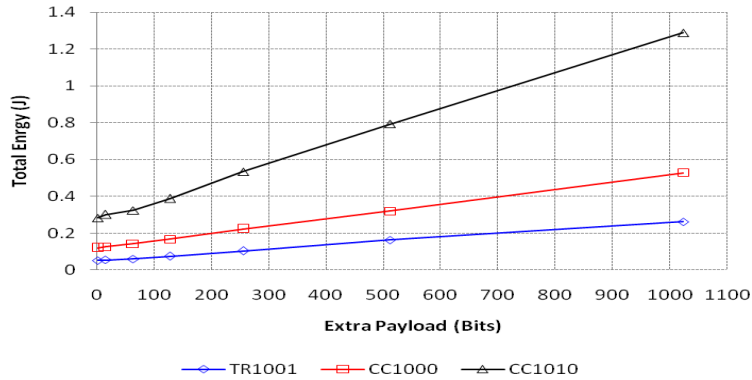

(a)

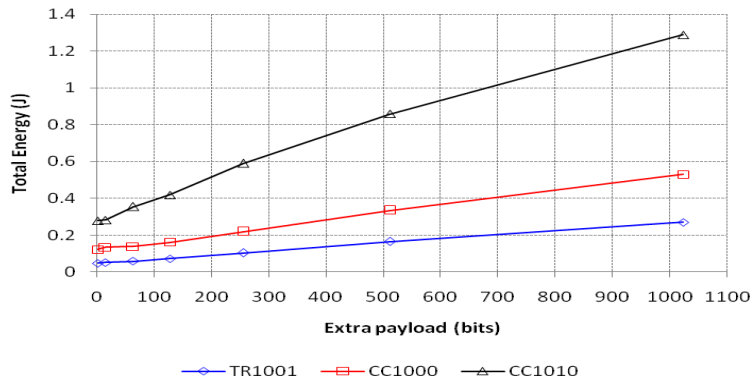

(b)

Figure 6: Comparison of Total Energy consumed by the chipsets TR1001, CC1000 and CC1010 for Leon (4x4) Crossed not shifted scheduling for (a) Random Deployment, (b) Grid deployment.

Figure 7 (a) and (b) shows a comparison of total energy consumed by the chipsets TR1001, CC1000 and CC1010 for Leon (4x4) Crossed shifted scheduling for (a) random deployment and (b) grid deployment. TR1001 consumes 0.262129 J and 0.270781 J of energy at the extra payload of 1023 bits for random and grid deployment respectively. It is also seen from that TR1001 consumes less energy in case of random deployment.

Figure 7 shows that CC1000 consumes an energy of 0.517003 J and 0.531041 J at extra payload of 1023 bits, for random and grid deployment respectively. It can be analyzed that CC1000 chipset consumes less energy in case of random deployment. The Figure 7 indicates that CC1010 consumes 1.272405 J and 1.306484 of energy for random and grid deployment respectively at extra payload of 1023 bits. It can be observed that energy consumed by CC1010 in case of random deployment is lower than grid deployment.

It is also observed from the Figure 7 that, Leon crossed shifted scheduling show a trivial difference in energy consumptions for the two types of deployment. Figure also indicates that TR1001 consumes minimum energy and CC1010 consumes maximum energy.

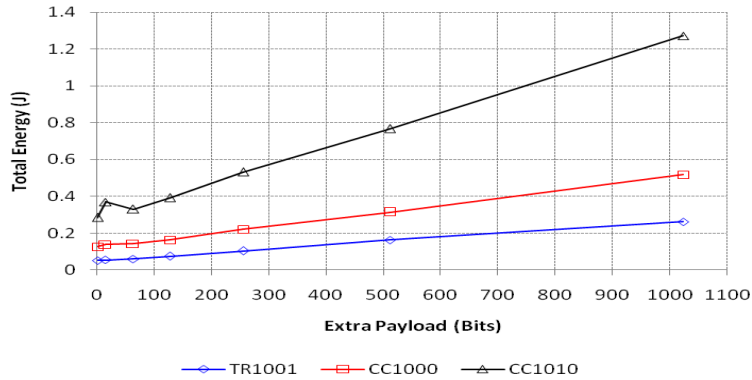

(a)

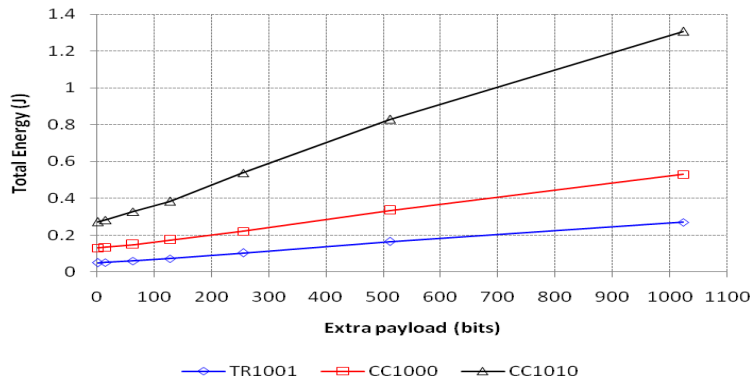

(b)

Figure 7: Comparison of Total Energy consumed by the chipsets TR1001, CC1000 and CC1010 for the Leon (4x4) Crossed Shifted scheduling for (a) Random Deployment, (b) Grid deployment.

Figure 8 (a) and (b) shows a comparison of total energy consumed by the chipsets TR1001, CC1000 and CC1010 for (4x4) Crossed Shifted for (a) random deployment and (b) grid deployment. It is observed that energy consumed by TR1001 is lowest among the other chipsets for both types of deployments. TR1001 consumes 0.250725 J and 0.270781J of energy at the extra payload of 1023 bits for random and grid deployment respectively. It is also concluded that TR1001 consumes less energy in case random deployment.

It can be observed from Figure 8 that CC1000 consumes an energy of 0.490086 J and 0.531859 J at extra payload of 1023 bits, for random and grid deployment respectively. It can be analyzed that CC1000 chipset consumes less energy in case of random deployment. The Figure 8 indicates that CC1010 consumes 1.189302 J and 1.284797 of energy for random and grid deployment respectively at extra payload of 1023 bits. It can be observed that energy consumed by CC1010 in case of random deployment is lower than grid deployment. It is indicated from the figures that 3x3 crossed shifted scheduling presents a minor difference in energy consumption for the three chips in both the random and grid scheduling.

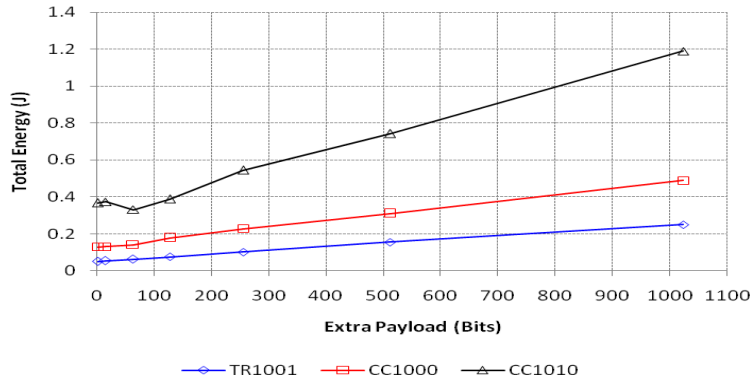

(a)

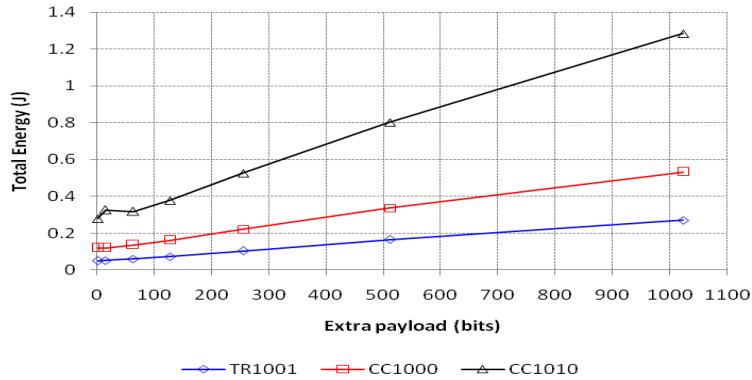

(b)

Figure 8: Comparison of Total Energy consumed by the chipsets TR1001, CC1000 and CC1010 for the (4x4) Crossed Shifted scheduling for (a) Random Deployment, (b) Grid deployment.

## 6. CONCLUSIONS

This paper focused on evaluating the amount of energy consumed during communication by few commercially available chipsets TR1001, CC1000 and CC1010. MERLIN protocol which integrates MAC and routing functions into the same architecture was used for the evaluation purpose. MERLIN is optimized for communication between nodes and gateways and supports upstream downstream and local broadcast types of traffic. Packet flow is achieved through a division of the network in time zones and through the usage of appropriate transmission scheduling. In particular, 6 scheduling tables, namely X, V, V (9x9), Leon (4x4) Crossed Shifted, Leon (4x4) Crossed not shifted, (4x4) Crossed shifted have been proposed and have been evaluated for the energy consumption. These scheduling tables are used in two different network scenarios one is where the nodes are deployed in a random fashion and another scenario where nodes were deployed in grid fashion.

The simulations presented results for extensive range and combination of parameters such as Extra payload, Scheduling tables, deployment types and chipsets. . It has been observed from the paper that for all the chipsets, the scheduling methods shows the same behavior of energy consumption. It is evident from the research that for all types of scheduling methods and deployment strategies, the chipset TR1001 consumes minimum energy among all the chipsets, and CC1010 consumes maximum energy. The chipsets CC1000 lies in between the two in terms of energy consumption. It has also been observed that V (9x9) and X scheduling methods shows a significant difference in energy consumption for both the deployment strategies while rest all methods shows minor difference in energy consumption for random and grid deployment strategies.

It can be concluded that energy consumption depends on the chipset, type of scheduling method used and the deployment strategy used for the deployment of the WSN. Hence a proper combination of these should be used in order to make an efficient use of energy.

Despite the encouraging results, we need to perform more experimentation to compare theses scheduling methods with some other WSN protocols except MERLIN. Future work will investigate and propose the new scheduling methods as well as new improved deployment strategies so as to reduce the energy consumption further.

**Authors**

**Monica** received her BE in Information Technology from the university of Jammu, J&K, India in 2007, M.Tech. in Computer Science and Engineering from National Institute of Technology, Jalandhar, Punjab,India in year 2009. Her M.Tech. Thesis was on "Simulative Investigations on Energy Consumption for Wireless Sensor Networks Using OMNeT++". She is presently working as Assistant professor in the Department of Computer Science and Engineering, National Institute of Technology, Jalandhar. Her major area of interest are Adhoc Networks and Wireless Sensor Networks.

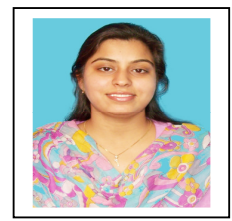

**Ajay K Sharma** received his BE in Electronics and Electrical Communication Engineering from Punjab University Chandigarh, India in 1986, MS in Electronics and Control from from Birla Institute of

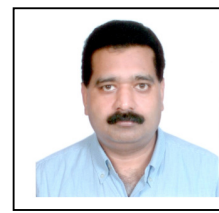

189

Technology (BITS), Pilani in the year 1994 and PhD in Electronics Communication and Computer Engineering in the yea r 1999. His PhD thesis was on "Studies on Broadband Optical Communication Systems and Networks". From 1986 to 1995 he worked with TTTI, DTE Chandigarh, Indian Railways New Delhi, SLIET Longowal and National Institute of technology (Erstwhile Regional Engineering College), Hamirpur HP at various academic and administrative positions. He has joined National Institute of Technology (Erstwhile Regional Engineering College) Jalandhar as Assistant Professor in the Department of Electronics and Communication Engineering in the year 1996. From November 2001, he has worked as Professor in the ECE department and presently he working as Professor in Computer Science & Engineering in the same institute. His major areas of interest are broadband optical wireless communication systems and networks, dispersion compensation, fiber nonlinearities, optical soliton transmission, WDM systems and networks, Radio-over-Fiber (RoF) and wireless sensor networks and computer communication. He has published 222 research papers in the International/National Journals/Conferences and 12 books. He has supervised 11 Ph.D. and 35 M.Tech theses. He has completed two R&D projects funded by Government of India and one project is ongoing. Presently he was associated to implement the World Bank project of 209 Million for Technical Education Quality Improvement programme of the institute. He is technical reviewer of reputed international journals like: Optical Engineering, Optics letters, Optics Communication, Digital Signal Processing. He has been appointed as member of technical Committee on Telecom under International Association of Science and Technology Development (IASTD) Canada for the term 2004-2007 and he is Life member of Indian Society for Technical Education (I.S.T.E.), New Delhi.